\documentclass[structabstract]{aa}  
\usepackage{longtable,lscape}
\usepackage{graphicx}
\usepackage{txfonts}

\usepackage{natbib}
\bibpunct{(}{)}{;}{a}{}{,} 

\begin{document}

   \title{Spectroscopy of brown dwarf candidates in IC~348 and the determination of its substellar IMF down to planetary masses\thanks{Based on observations obtained with WIRCam, a joint project of CFHT, Taiwan, Korea, Canada, France, and MegaPrime/MegaCam, a joint project of CFHT and CEA/DAPNIA, at the Canada-France-Hawaii Telescope (CFHT) which is operated by the National Research Council (NRC) of Canada, the Institute National des Sciences de l'Univers of the Centre National de la Recherche Scientifique of France, and the University of Hawaii.}
}

\author{C. Alves de Oliveira\inst{1}, E. Moraux\inst{2}, J. Bouvier\inst{2}, G. Duch\^ene\inst{3, 2}, H. Bouy\inst{4}, T. Maschberger\inst{2}, P. Hudelot\inst{5}}

\institute{European Space Astronomy Centre (ESA), P.O. Box 78, 28691 Villanueva de la Ca\~{n}ada, Madrid, Spain\\
\email{calves@sciops.esa.int}
\and UJF-Grenoble 1/CNRS-INSU, Institut de Plan\'etologie et d'Astrophysique de Grenoble (IPAG) UMR5274, Grenoble, 38041, France
\and Astronomy Department, University of California, Berkeley, CA 94720--3411, USA
\and Centro de Astrobiolog'a (INTA-CSIC); LAEFF, P.O. Box 78, 28691 Villanueva de la Ca\~{n}ada, Spain
\and Institut d$^{\prime}$Astrophysique de Paris, UMR 7095 CNRS, Universit\'e Pierre et Marie Curie, 98bis Bd Arago, 75014 Paris, France}
\date{Received 14 August 2012; accepted 8 November 2012}

   \abstract
   {Brown dwarfs represent a sizable fraction of the stellar content of our Galaxy and populate the transition between the stellar and planetary mass regime. There is however no agreement on the processes responsible for their formation. }
   {We have conducted a large survey of the young, nearby cluster IC~348, to uncover its low-mass brown dwarf population and study the cluster properties in the substellar regime.}    
  {Deep optical and near-IR images taken with MegaCam and WIRCam at the Canada-France-Hawaii Telescope (CFHT) were used to select photometric candidate members. A spectroscopic follow-up of a large fraction of the candidates was conducted to assess their youth and membership. }
  {We confirmed spectroscopically 16 new members of the IC~348 cluster, including 13 brown dwarfs, contributing significantly to the substellar census of the cluster, where only 30 brown dwarfs were previously known. Five of the new members have a L0 spectral type, the latest-type objects found to date in this cluster. At 3~Myr, evolutionary models estimate these brown dwarfs to have a mass of $\sim$13~M$_{\emph{Jup}}$. Combining the new members with previous census of the cluster, we constructed the IMF complete down to 13~M$_{\emph{Jup}}$.}
  {The IMF of IC~348 is well fitted by a log-normal function, and we do not see evidence for variations of the mass function down to planetary masses when compared to other young clusters.}
 
\titlerunning{Spectroscopy of brown dwarf candidates in IC~348} 
\authorrunning{C. Alves de Oliveira et al.} 

   \maketitle


\section{Introduction}
\label{intro}

Large scale imaging surveys of young star-forming clusters and spectroscopic confirmation of bona-fide members are fundamental to derive complete and reliable census of their populations. Such endeavors are the first step in the characterization of a cluster, allowing the posterior statistical study of the members' properties (e.g. disks and jets, dynamics, binarity, or the mass distribution) and comparison between different clusters. In the last decade, several cluster's in our Galaxy have been extensively studied with complementary techniques \citep{Reipurth2008a,Reipurth2008b}, though often these are not sensitive to the lowest mass members. 

The initial mass function \citep[IMF, ][]{Salpeter1955,Scalo1986,Kroupa2002,Chabrier2003} is one of the main properties studied in clusters. The current knowledge from observations and efforts of theoretical work to understand its origin have been recently summarized, for example, by \citet{Bastian2010}, \citet{Kroupa2011}, and \citet{Jeffries2012}. In the substellar regime, the characterization of the mass spectrum of a cluster gains another importance, since it could help distinguish among different formation scenarios for brown dwarfs \citep[e.g.,][]{Hennebelle2012}. This is the goal of a large program (P.I. J. Bouvier) we are conducting at the Canada--France--Hawaii Telescope (CFHT), designed to uncover the substellar population of various nearby young cluster and compare their IMF and minimum mass for brown dwarf formation \citep[][]{Burgess2009,AlvesdeOliveira2010,AlvesdeOliveira2012,Spezzi2012}.

In this paper, we present the results of this survey for IC~348, where we aim at uncovering its substellar population down to the planetary regime. The properties known to date of this nearby young cluster and its population have been summarized, for example, in \citet{Herbst2008} and \citet{Stelzer2012}. In short, IC~348 is located at 300~pc and is part of the OB Per association, has an estimated age of 2-3~Myr, and the known population amounts to 345 members, including 30 brown dwarfs. The starting point of our survey are new deep optical and near-IR photometric observations obtained at the CFHT (described in Sect.~\ref{obs}) to uncover low-mass brown dwarf candidates that have evaded detection in previous surveys (Sect.~\ref{sel}), either because these studies were shallower or did not observe the outskirts of the cluster with the required sensitivity. Through spectroscopically follow-up, we derive a clean and complete census of the substellar population (Sect.~\ref{mem}), and extend the IMF down to the planetary regime (Sect.~\ref{sectionimf}). 


\section{Observations and data reduction} \label{obs}
In this section, we describe the various data sets collected for this study.

\subsection{The CFHT Megacam \& WIRCam photometric surveys}
An optical and near-IR photometric observational campaign of IC~348 was conducted at the CFHT using MegaCam (programme 06BF28), an optical camera with a wide field of 1$^{\circ}$~$\times$~1$^{\circ}$ \citep{Boulade2003}, and WIRCam (programme 06BF23), a near-IR camera covering 20$\arcmin$~$\times$~20$\arcmin$ \citep{Puget2004}. The optical observations consisted of one field observed with the \emph{z$^{\prime}$} filter and three different individual exposure times and number of integrations: 300~$s$~$\times$~5 (dithering)~$\times$~6 (micro-dithering), 30~$s$~$\times$~5~(dithering), and 3~$s$~$\times$~5~(dithering). The same observing strategy was used in the near-IR, where for each filter two images were taken with a short and a long exposure time. Short exposures were defined as 5~$s$~$\times$~7~(dithering) for the three bands. For the long exposures, the following times were used: 45~$s$~$\times$~7~(dithering)~$\times$~4~(micro-dithering) for \emph{J}, 10~$s$~$\times$~7~(dithering)~$\times$~8~(micro-dithering) for \emph{H}, and 15~$s$~$\times$~7~(dithering)~$\times$~4~(micro-dithering) for \emph{K}$_{s}$. In our previous paper on the search for young T~dwarfs in IC~348, \citet{Burgess2009} presented briefly the optical and near-IR observations taken with long exposures. For this study, we have re-analysed all the images in a consistent way, tailoring the photometric extraction to our purpose of collecting the most complete set of reliable detections. To that end, we include also the short exposure \emph{z$^{\prime}$JHK}$_{s}$ observations, which allow us to extend the range of magnitudes we are probing. The MegaCam observations were taken with a single pointing, while four different WIRCam pointings were needed to cover the cluster. The pointing coordinates of the observed fields are presented in Table~1 of \citet{Burgess2009}, and are the same for short and long observations. Given that the MegaCam field encompasses in its totality the smaller area WIRCam mosaic, we restrict our analysis to the overlapping region (34\arcmin~x~34\arcmin) and hereafter refer to it as the `CFHT field'.

All data were reduced following the steps described in our previous survey papers \citep[see, for example,][]{Burgess2009,AlvesdeOliveira2010}. Summarily, individual images were first processed by the 'I'iwi reductions pipeline at the CFHT, which includes detrending (e.g. bias subtraction, flat-fielding, non-linearity correction, cross-talk removal), sky subtraction, and preliminary astrometric calibration. The final astrometric and photometric calibrations, as well as the production of the stacked mosaics were handled at TERAPIX \citep{2007ASPC..376..285M}. 
\begin{figure}
\centering
\includegraphics[width=\columnwidth]{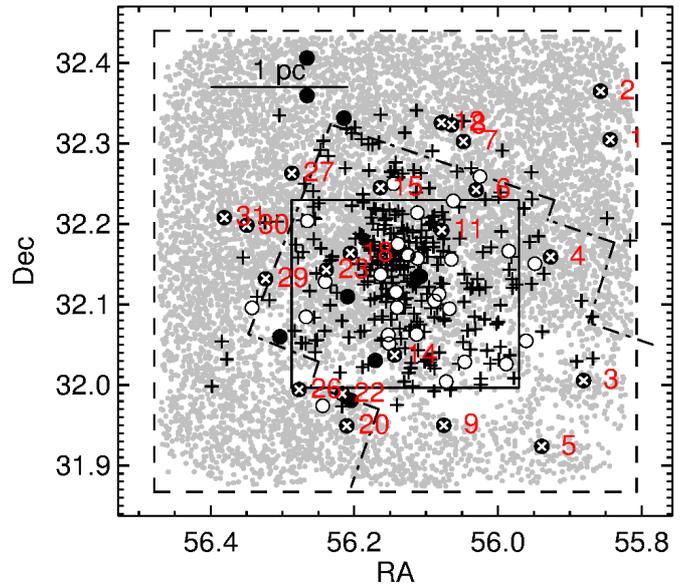}
\caption{Plotted over the entire CFHT catalogue for IC~348 (grey dots) are the candidate substellar members selected in this study (black filled circles, white crosses signal spectroscopic observations, numbers correspond to the running identification), and all the candidate members of the cluster from the literature present in the CFHT catalogue (plus sign), including the previously known brown dwarfs from the literature (open circles). The dashed line represents the field observed with WIRCam, the dash-dot line the X-ray observations analysed in \citet{Stelzer2012}, and the solid line delineates the area for which there is a complete census of the cluster for masses $\ge$~30~M$_{\emph{Jup}}$ \citep{Luhman2003ic348} .}
\label{sky}
\end{figure}

Photometric measurements were done using PSFEx and SExtractor tools\footnote{Available at http://www.astromatic.net/} \citep{2011ASPC..442..435B,1996A&AS..117..393B}. Firstly, PSFEx selects point-sources with well-defined stellar profiles and computes a PSF model capable of representing smooth PSF variations across large mosaics. Secondly, the PSF model is fed into SExtractor which detects and analyses the sources in the astronomical images in an automated way. The extraction process was repeated two times for each near-IR image, in one case requiring that an object is extracted if the flux measured is 5~$\sigma$ above the estimated background (to ensure the detection of sources near bright stars), and in the other relaxing the detection threshold to 3~$\sigma$. We chose this approach in order to recover the faint objects detected \emph{by eye}, while minimizing spurious detections. For each image, the 3 and 5~$\sigma$ catalogues were merged, keeping the higher signal--to--noise measurements for duplicate detections. We merged the resulting catalogues into two lists (short and long exposures) that contain only objects detected across all 3 \emph{JHK} bands. A single catalogue was then created by cross-matching these two, and requiring a positional match of $\lesssim$1$\arcsec$. In the overlapping magnitude range for a single filter, we kept the measurements that had the smaller photometric errors in the combined \emph{JHK$_{s}$} photometry. The analysis for the \emph{z$^{\prime}$} band was done in a similar fashion, where catalogues were extracted for the three images with different integration times, and a single list was produced by merging those, keeping the best photometric detections for duplicates. The optical data was similarly merged with the near-IR catalogue. The near-IR catalogue contains 10631 sources with a detection in the \emph{JHK$_{s}$} bands, of which 75\% are also detected in the \emph{z$^{\prime}$} band. Completeness limits are defined by the long exposure catalogues and were already derived by \citet{Burgess2009} to be $\sim$23.5 (\emph{z$^{\prime}$}), 21.5 (\emph{J}), 20.0 (\emph{H}), and 18.9 (\emph{K$_{s}$}). 

The photometric calibration of the WIRCam data is done with 2MASS stars in the observed frames as part of the nominal pipeline reduction, with a typical error on the zero-point of $\sim$0.05~mag. Colour corrections are not applied, and therefore the photometry presented here is given in the CFHT Vega system. We compared the magnitudes of the short exposures to the 2MASS point-source catalogue and found the mean magnitude differences between the two systems to be 0.06, 0.03, and 0.07 mag for \emph{JHK$_{s}$} respectively, which are of the order of the zero-point uncertainties. The mean differences between the short and long WIRCam exposures is of 0.07, 0.01, and 0.009 for \emph{JHK$_{s}$}, respectively. The photometric calibration of the MegaCam data is done using standard stars routinely observed by the queue service observing team at the CFHT. We compared the final MegaCam \emph{z$^{\prime}$}-band photometry with data from the Wide Field Camera \citep[WFCAM,][]{Casali2007} at the UKIRT telescope, that observed this cluster as part of the UKIRT Infrared Deep Sky Survey \citep[UKIDSS,][]{Lawrence2007}, and found a mean magnitude difference between the two systems of 0.04~mag.

\subsection{The Osiris / GranTeCan spectroscopic follow-up}

We conducted a spectroscopic follow-up for a subsample of 13 candidate members of IC~348, with optical magnitudes brighter than $\sim$20~mag in \emph{z$^{\prime}$}. The selection of  candidate members is described in Sect.~\ref{select:cmd}. All targets were observed with the OSIRIS spectrograph \citep{Cepa2000} on the GTC (Gran Telescopio Canarias, La Palma, Canary Islands), in queue-service mode. The spectra were taken with the R300R grism (R$\sim$300) and an 1$\arcsec$ slit aligned with the parallactic angle, following a nodding pattern for posterior sky-subtraction. The dates and exposures times for each target are shown in Table~\ref{log1}. The spectro-photometric standards, flats, bias, and arcs necessary for the calibrations were also observed in the same nights. We reduced the data with IRAF standard routines. Briefly, we applied bias subtraction and flat--field correction to the 2D spectra, and subtracted adjacent pairs of spectra to subtract the sky contribution. The 1D spectra were extracted and wavelength calibrated. The instrumental response was corrected with the observed spectrophotometric standards, and the individual spectra were median-combined. 
\begin{table}
\caption{Journal of the spectroscopic observations.}            
\begin{tabular}{l l l }        
\hline \hline
Target\tablefootmark{a} & Date &  Exp. Time \\
   \hline                        
 \multicolumn{3}{c}{OSIRIS/GTC} \\
  \hline  
CFHT-IC348-2 & 20-09-2010  &  2$\times$300\emph{s}  \\      
CFHT-IC348-3 & 20-09-2010  &  2$\times$300\emph{s}  \\     
CFHT-IC348-5   & 07-09-2010  &  2$\times$300\emph{s}   \\    
CFHT-IC348-6 & 24-12-2010  &  2$\times$1800\emph{s}   \\    
CFHT-IC348-7 & 05-09-2010  &  2$\times$150\emph{s}   \\     
CFHT-IC348-8 & 05-09-2010  &  2$\times$150\emph{s}   \\     
CFHT-IC348-9 & 05-09-2010  &  2$\times$150\emph{s}   \\     
CFHT-IC348-12  & 02-10-2010  &  2$\times$600\emph{s}   \\    
CFHT-IC348-22 & 28-10-2010  &  2$\times$2100\emph{s}  \\     
CFHT-IC348-23 & 27-10-2010  &  2$\times$2100\emph{s}  \\     
CFHT-IC348-26  & 21-09-2010  &  2$\times$600\emph{s}   \\   
CFHT-IC348-30   & 08-09-2010  &  2$\times$300\emph{s}   \\   
CFHT-IC348-31  &  20-09-2010  &  2$\times$600\emph{s}   \\     
   \hline                        
 \multicolumn{3}{c}{GNIRS/Gemini} \\
  \hline  

CFHT-IC348-1	&	   07-10-2011	   &	   12$\times$300\emph{s}	\\	 
CFHT-IC348-4	&	   17-09-2011	   &	   12$\times$300\emph{s}	\\	  
CFHT-IC348-6	&	   15-08-2011	   &	   12$\times$240\emph{s}	\\    	  
CFHT-IC348-8	&	   04-01-2011	   &	   12$\times$240\emph{s}	\\      
CFHT-IC348-10	&	   21-09-2011	   &	   14$\times$300\emph{s}	\\	  
CFHT-IC348-11	&	   01-09-2011	   &	   6$\times$300\emph{s}	\\	 
CFHT-IC348-13 	&	   05-10-2011	   &	   8$\times$300\emph{s}	\\	 
CFHT-IC348-14	&	   13-10-2011	   &	   12$\times$300\emph{s}	\\	 
CFHT-IC348-15	&	   21-08-2011	   &	   8$\times$300\emph{s}	\\	 	
CFHT-IC348-18	&	   17-09-2011	   &	   12$\times$300\emph{s}	\\ 	  
CFHT-IC348-20	&	   01-09-2011	   &	   12$\times$240\emph{s}	\\	 			
CFHT-IC348-22	&	   31-12-2010	   &	   12$\times$240\emph{s}	\\	 		
CFHT-IC348-23	&	   15-08-2011	   &	   12$\times$240\emph{s}	\\	 	
CFHT-IC348-27	&	   17-09-2011	   &	   12$\times$300\emph{s}	\\	 	
CFHT-IC348-29	&	   20-09-2011	   &	   10$\times$300\emph{s}	\\	 
2MASS~J04373705+2331080\tablefootmark{b}	&	   10-10-2011	   &	   12$\times$300\emph{s}	\\	 
\hline   
\end{tabular}
\label{log1} 
\tablefoottext{a}{Coordinates for the CFHT-IC348 targets are given in Table~\ref{candidates}.} 
 
\tablefoottext{b}{Brown dwarf member of Taurus (2MASS~J04373705+2331080, 1~Myr), classified in the optical as an L0 \citep{Luhman2009}, and used as empirical template for spectral classification.} 

\end{table}

\subsection{The GNIRS / Gemini spectroscopic follow-up}
We observed in the near-IR the candidates that were too faint for optical spectroscopy. These amount to 15 targets (4 of which were also observed with Osiris/GTC) with magnitudes fainter than $\sim$17 in the \emph{J} band. Based on their photometric properties and evolutionary models (see Sect.~\ref{sel}), the new candidate brown dwarfs observed could have spectral types in the L-type regime. Only a few L-type brown dwarfs have been spectroscopically confirmed in nearby young clusters (1-3~Myr), which is the reason why we use young-field L-dwarfs \citep[10-30~Myr,][]{Cruz2009} for late-type spectral classification. However, to validate our fitting method in the L-type spectral regime, we additionally observed spectroscopically with same instrumental configuration a previously known young brown dwarf member of Taurus (2MASS~J04373705+2331080, 1~Myr), classified in the optical as an L0 \citep{Luhman2009}. 

All targets were observed with the GNIRS spectrograph \citep{Elias2006}, mounted on the Gemini North (GN) telescope. Two targets were observed in December 2010 and January 2011 during the instrument's science verification (programme GN-2010B-Q-47). The other targets were observed in queue-schedule observing mode from August to October 2011 (programme GN-2011B-Q-8). Observations were taken in cross-dispersed mode, with the short (0.15 arcsec/pix) blue camera, the 32 line~mm$^{-1}$ grating (R$\sim$1000), and a 0.8$\arcsec$ slit. Consecutive exposures were taken with an ABBA pattern. The dates of observations, individual exposure times and number of integrations are given in Table~\ref{log1}. Standard stars were observed before or after the target at a similar airmass.

The data were reduced with the Gemini IRAF package (v1.11) provided by the Gemini Observatory. Firstly, we cut into extensions the different orders of the calibrations files (flat fields, arcs, and pinholes), and created the flat field for each order. Secondly, the science spectra were cut, flatfielded, and sky-subtracted, and finally median combined. The pinhole and arc spectra were used to determine the spectral distortion and wavelength calibration, which were applied to the science data before extracting the final spectra. Telluric line removal was done by dividing the target's spectra by those of the corresponding standard stars. The relative flux was recovered by multiplying the resulting spectra by a theoretical spectrum of the same temperature as the standard stars \citep{Pickles1998}. 

Targets 10 and 13 did not have enough signal-to-noise and could not be reduce properly, and are therefore not included in the spectral analysis. They remain however, photometric candidate members.

   \begin{figure*}
   \centering
 \includegraphics{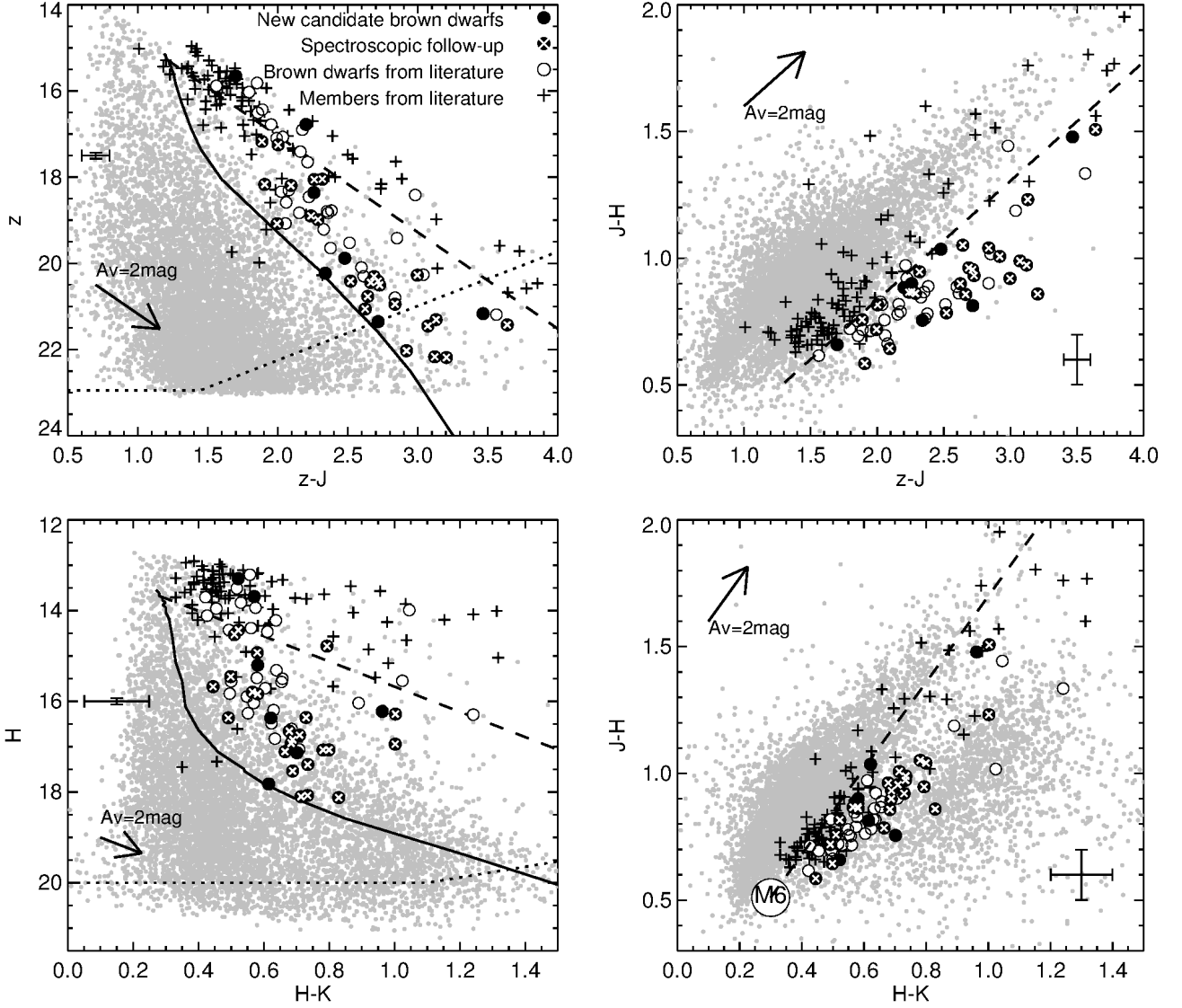}
   \caption{Colour-magnitude (CMD) and colour-colour diagrams for the IC~348 cluster. In the CMDs, the solid line represents the DUSTY 3~Myr isochrone \protect{\citep{Chabrier2000}}, and the dashed lines the $\sim$75~M$_{\emph{Jup}}$ limit, with increasing amount of visual extinction. The dotted line shows the photometric completeness limits of the CFHT MegaCam and WIRCam survey. Symbols are the same as in Fig.~\ref{sky}. Typical error bars are shown in all diagrams.}
   \label{cmd}
    \end{figure*}
   \begin{figure*}
   \centering
 \includegraphics{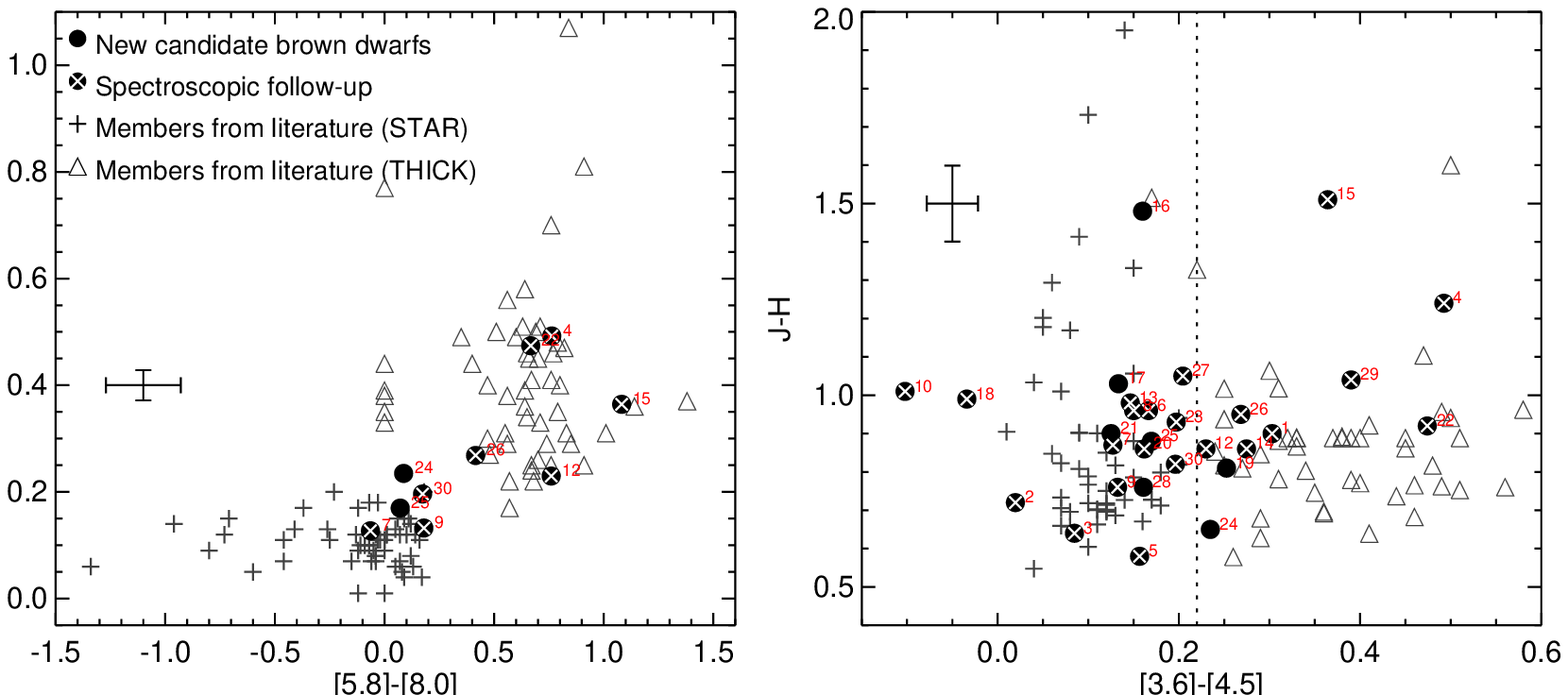}
   \caption{IRAC / \emph{Spitzer}  (\emph{left-hand-side}), and WIRCam/CFHT and IRAC/\emph{Spitzer} (\emph{right-hand-side}) colour-colour diagrams. New candidate members (filled circles) are compared with previously known members classified from the IRAC SED slope as being diskless (\emph{plus sign}, `STAR') or surrounded by thick disks (\emph{triangle}, `THICK').}
   \label{spitzer}
    \end{figure*}

\subsection{Mid--IR: \emph{Spitzer} space telescope and WISE}
\label{data:spitzer}
To complement the CFHT data set on IC~348, we included in this study archival mid-IR data to better assess the likelihood of membership for the candidate members and characterize their properties. Infrared excess around young stellar objects (YSOs) provides evidence for discs and its detection is commonly used as an indicator of youth \citep{Haisch2001}. We have merged the CFHT catalogue with data from the \emph{Spitzer} space telescope and the Wide-field Infrared Survey Explorer \citep[WISE,][]{Wright2010}.
 
The Perseus molecular cloud, including the IC~348 cluster, has been mapped with \emph{Spitzer}'s Infrared Camera \citep[IRAC]{Fazio2004} in the 3.6, 4.5, 5.8 and 8.0~$\mu$m bands over a region of $\sim$3.86~deg$^{2}$ \citep{Jorgensen2006} and with the Multiband Imaging Camera \citep[MIPS]{Rieke2004} in the 24 and 70~$\mu$m bands over a total of $\sim$10.5~deg$^{2}$ \citep{Rebull2007}, which encompass the CFHT field in its totality. The data were retrieved from the C2D point-source catalogues of the final data delivery \citep{Evans2005}. All fluxes were converted to magnitudes using the following zero-points from the instruments' handbooks: 280.9$\pm$4.1, 179.7$\pm$2.6, 115.0$\pm$1.7, 64.1$\pm$0.94~(Jy), for the 3.6, 4.5, 5.8 and 8.0~$\mu$m IRAC bands, respectively, and 7.17$\pm$0.11~(Jy) for the 24~$\mu$m MIPS band. From our candidate members (Sect.~\ref{sel}), no target was detected at 70~$\mu$m. The \emph{Spitzer} catalogues were merged with the CFHT detections catalogue, requiring the closest match to be within 1$\arcsec$. A counterpart was found for 67$\%$ objects that were detected in one or more mid-IR bands. We have also crossed-matched the CFHT catalogue with the WISE preliminary release of its all-sky source catalogue, which includes observations taken at 3.4, 4.6, 12 and 22~$\mu$m with positions reconstructed using the 2MASS point source catalogue. For the IC~348 cluster, the WISE observations are shallower then \emph{Spitzer}'s at all wavelengths, and we did not include them further in the analysis since they did not result in any additional detection for our candidate members.


\section{Selection of new candidate brown dwarfs}\label{sel}
\label{select:cmd}

\subsection{Colour$-$colour and colour$-$magnitude diagrams}

We follow the same selection criteria adopted in \citet{AlvesdeOliveira2010} to identify candidate brown dwarfs members of the cluster. We define as candidates objects that have colours redwards from the 3~Myr Dusty isochrone \citep{Chabrier2000} in the colour-magnitude diagrams of all combinations of filters (\emph{z$^{\prime}$} vs. \emph{z$^{\prime}$}-\emph{J},  \emph{z$^{\prime}$} vs. \emph{z$^{\prime}$}-\emph{H}, \emph{z$^{\prime}$} vs. \emph{z$^{\prime}$}-\emph{K$_{s}$}, \emph{J} vs. \emph{J}-\emph{H},  \emph{J} vs. \emph{J}-\emph{K$_{s}$}, and \emph{H$_{s}$} vs. \emph{H}-\emph{K$_{s}$}). We take 300~pc to be the distance to the cluster. In the near-IR colour-colour diagram, we further constrain the sample to objects that have colours redder than the empirical photospheric colours of a young M6 dwarf \citep{Luhman2010} in the WIRCam system \citep{AlvesdeOliveira2012}, progressively reddened by extinction \citep{Rieke1985}. In the \emph{J}-\emph{H} vs. \emph{z$^{\prime}$}-\emph{J} colour-colour diagram, we define an empirical division along the extinction vector based on the separation between stellar and substelar members of the cluster from the literature.

The IC~348 cluster has been throughly studied and the most significant compilations of members were done by \citet{Luhman2003ic348,Luhman2005} and \citet{Muench2007}, totaling 345 members confirmed spectroscopically. We cross-matched this list with our catalogues and recover 205 members of the cluster. From the remaining members, 138 are saturated in our images. One object (2MASS~J03440577+3200284) has been classified as a Class~I member by \citet{Muench2007} from mid-IR observations but is not detected in our \emph{J} band therefore being excluded from the merged catalogues. We only miss 1 object that should be in the final catalogue (2MASS~J03442186+3217273), it is close to a saturated star and was not detected by the extraction algorithm. Taking into account the photometric errors, the criteria we defined to select substellar candidate members, recovers 100\% of the cluster's known brown dwarfs.

Figures~\ref{sky} and \ref{cmd} show all the detections in the CFHT catalogue (light grey dots), the known population of the cluster (plus sign) including the previously confirmed brown dwarfs (open circles), and the new candidate brown dwarfs (filled black circles) including the ones observed spectroscopically (white crosses). All candidates were inspected visually in the images taken with the four filters, and we removed the ones that were contaminated by artifacts or bad pixels, and likely not to be real young members. We removed from the candidate list any object that was already spectroscopically classified as a member of the cluster, and also confirmed that none of the candidates had been previously classified as a contaminant \citep[e.g.,][]{Luhman2003ic348}. The final list of new candidate brown dwarfs contains 31 objects, which are listed in Table~\ref{candidates}. We have conducted a spectroscopic follow-up of 24 of these candidates. 

\subsection{Mid-IR photometric properties of candidate members}

Young stellar objects can be surrounded by disks or dusty envelops, remnants from their formation process, observed as an excess of mid-IR emission with respect to the photosphere. From the 31 candidate substellar members, only 10 have reliable photometry in all IRAC bands in the \emph{Spitzer} catalogue, and all have only upper limits at the 24~$\mu$m band. In the IRAC colour-colour diagram in Fig.~\ref{spitzer}, we plot the previously known members studied by \citet{Lada2006}, which were classified from the IRAC spectral energy distribution (SED) slope as being diskless (plus sign, `STAR') or surrounded by thick disks (triangle, `THICK'). Five of our candidate members (CFHT-IC348-4, 12, 15, 22, and 26) show evidence for mid-IR excess (marked as \emph{ex1} in Table~\ref{candidates}). Five candidates have colours consistent with stellar photospheres (CFHT-IC348-7, 9, 24, 25, 30). 

Given that all but one candidate are detected at 3.6 and 4.5~$\mu$m, though not in the other bands, we combine the IRAC data at these bands with the WIRCam photometry to look for excesses in the other candidates. In the right panel of Fig.~\ref{spitzer}, the two groups of the known members show a clear separation, and we use this empirical division ([3.6]-[4.5]$\sim$0.2~mag) to classify 9 of our candidates (CFHT-IC348-1, 4, 12, 14, 15, 19, 22, 26, 29) as having excess at short mid-IR wavelengths (marked as \emph{ex2} in Table~\ref{candidates}). The target CFHT-IC348-24 is only marginally redwards of the empirical division, and in the IRAC colour-colour diagram has colours consistent with a stellar photosphere, and therefore it is not classified as having IR excess. Combining the results from both diagrams, we find that nearly one third of the candidate members show evidence for mid-IR excess. 


\section{Spectroscopic follow-up}\label{mem}

\subsection{Numerical spectral fitting}
We apply the numerical fitting technique described in \citet{AlvesdeOliveira2010,AlvesdeOliveira2012} to each target to derive spectral types and visual extinction (A$_{V}$). Summarily, the fitting method relies on minimizing a goodness-of-fit statistic between the candidate and an empirical grid of young templates to determine the best fit for both parameters. The GNIRS/Gemini targets are compared to a grid consisting of near-IR spectra of members of Taurus, Chamaeleon, and IC~348 (1-3~Myr), with spectral types determined in the optical through comparison to averages of dwarfs and giants, varying from M3 to M9.5 \citep{Briceno2002,Luhman2003taurus,Luhman2003ic348,Luhman2004}, and near-IR spectra (K. Luhman, priv. comm.) of optical standard young-field L dwarfs \citep{Cruz2009} with spectral types from L0 to L5 and spectral features indicative of low gravity. The Osiris/GTC targets are compared to a grid consisting of the optical spectra of the same targets and additional members of Chamaeleon I (1~Myr) with known spectral types determined in an analogous way \citep{Luhman2004cha}. As shown in \citet{AlvesdeOliveira2012}, the uncertainties of the fitting method are of the order of 1 spectral subclass and 1 magnitude of visual extinction, though they can be larger if the signal-to-noise of the target is low, or the target has a disk seen at a particular geometry, e.g., edge-on.

   \begin{figure}
   \centering
 \includegraphics[width=\columnwidth]{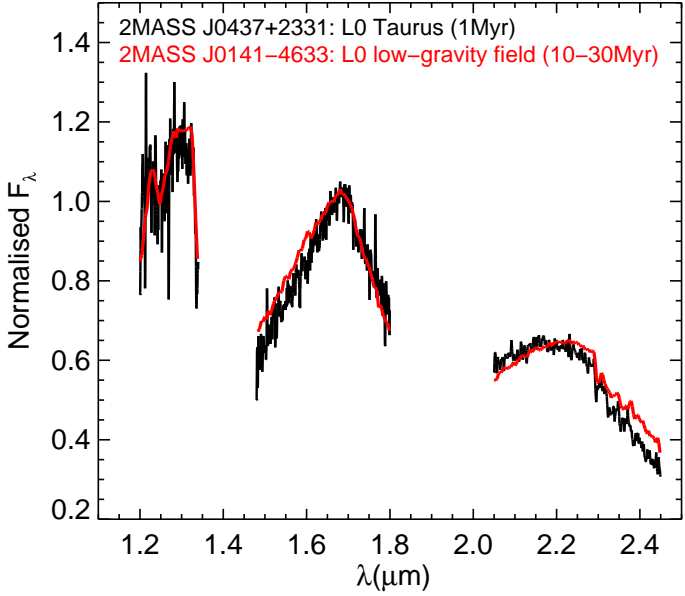}
   \caption{GNIRS/Gemini spectrum of 2MASS~J04373705+2331080, a young L0 member of Taurus confirmed spectroscopically in the optical \citep{Luhman2009}. The best-fitting template, 2MASS~J01415823$-$4633574, young-field L0 dwarf \citep{Cruz2009}, is shown in red.}
   \label{l0taurus}
    \end{figure}

To further validate our fitting method, we applied it to the young brown dwarf member of Taurus (2MASS~J04373705+2331080), and found the best fit to correspond to that of an L0, that in our grid corresponds to 2MASS~J01415823$-$4633574  \citep[10-30~Myr][]{Cruz2009}, and an A$_{V}$$=$3.3~mag (Fig.~\ref{l0taurus}). This result adds up to the tests already presented in \citet{AlvesdeOliveira2012} in showing the capabilities of the fitting method. The value derived for visual extinction, is significantly higher than the zero value reported in \citet{Luhman2009}. However, if we estimate the A$_{V}$ from the near-IR \emph{J}-\emph{H} vs. \emph{H}-\emph{K$_{s}$} diagram, by dereddening the colours of the object along the extinction vector till they match the expected colours for young L0 dwarfs \citep{Luhman2010}, we find an A$_{V}$$=$2.1~mag. This is consistent, within the errors, to the value found through the spectra fitting. Additionally, the differences between the shape of the \emph{K}-band part of the spectra, which are indicative of the low gravity of the Taurus brown dwarf, could introduce an additional source of error in the determination of the extinction.

We applied the numerical spectral fitting procedure to the 22 candidates and derived spectral types from M5.5 to L0 in the optical spectra, and from M7 to L0 in the near-IR spectra. Four targets have been observed at both wavelength ranges, these are CFHT-IC348-6, 8, 22, and 23. The difference in classification between optical and near-IR is 1.75, $-$1.25, $-$0.5, and 1.5 spectral classes, respectively. Since the spectral fitting method has an estimated error of $\sim$1 spectral type for both regimes, we find these results to be in good agreement. Also, there is no clear trend in the differences, as was already found in the tests presented in \citet{AlvesdeOliveira2012}. The differences in the extinction values are of $-$2, 0.3, 1.2, and 5.2 visual magnitudes. Only the last target (CFHT-IC348-23) shows a large discrepancy in the value for extinction. When we dereddened its near-IR colours along the extinction vector till they match the colours expected by either the optical or near-IR spectral type (M8.5 or M7, respectively), we find values of 3~mag and 4~mag. These are consistent with the value derived from the near-IR spectrum and it indicates that the low signal-to-noise of the optical spectrum could be compromising the reliability of the fit.

\begin{figure}
   \centering
 \includegraphics[width=\columnwidth]{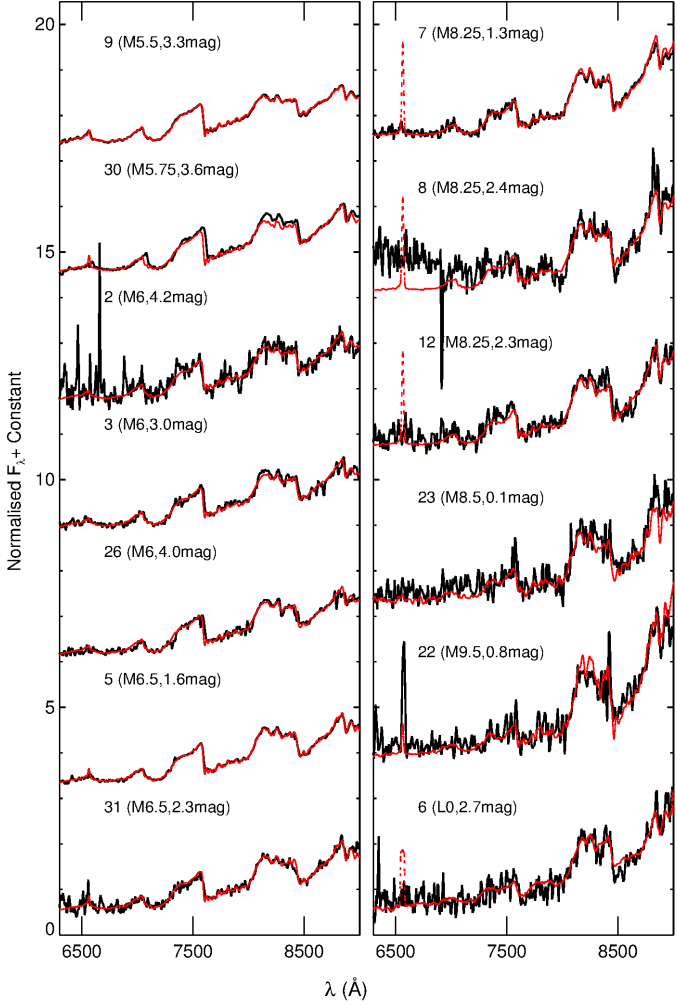}
   \caption{Osiris/GTC optical spectra of candidate members in IC~348. All spectra are corrected for extinction with the values found through the numerical fitting procedure. The best-fitting template is overplotted in red.}
   \label{optspec}
    \end{figure}

\begin{figure}
   \centering
 \includegraphics[width=\columnwidth]{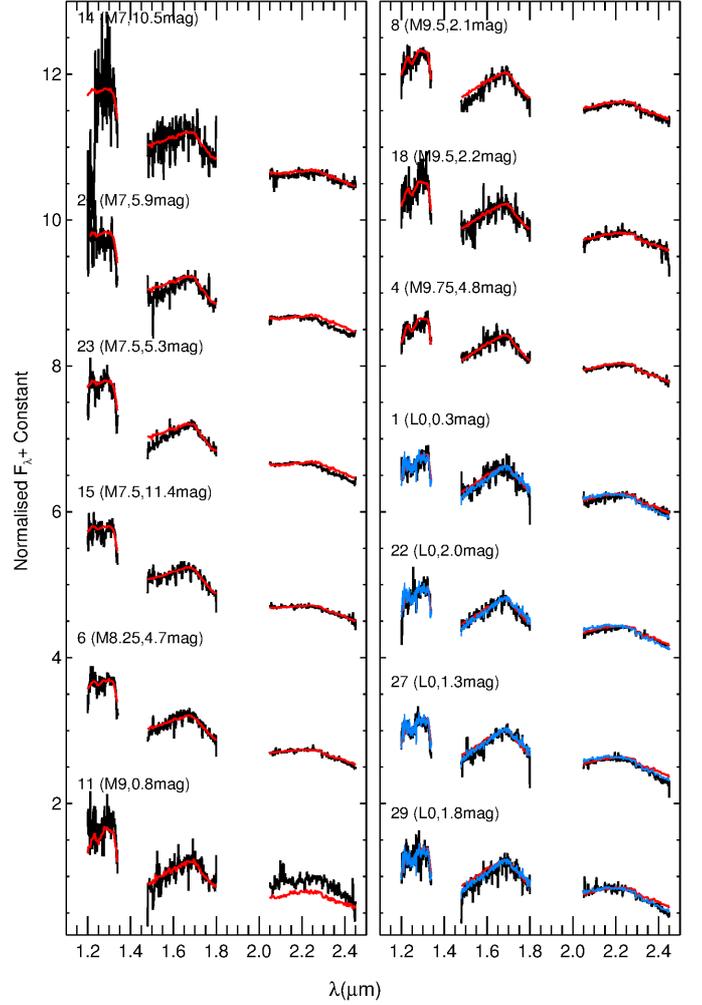}
   \caption{GNIRS/Gemini near-IR spectra of candidate members in IC~348. All spectra are corrected for extinction with the values found through the numerical fitting procedure. The best-fitting template is overplotted in red. The spectrum of the young L0 in Taurus, 2MASS~J01415823$-$4633574, is overplotted in blue for the L0 candidates.}
   \label{irspec}
    \end{figure}

The results of the numerical fitting method are presented in Table~\ref{candidates}. Figures~\ref{optspec} and \ref{irspec} show the dereddened spectra, with the best-fitting template overplotted in red. Four targets with near-IR spectra are classified as L0, and for those we also overplot the spectrum of 2MASS~J04373705+2331080, for comparison. 

\subsection{Membership}
To confirm the membership of each target observed spectroscopically, we look for signatures of low-gravity or/and accretion discs characteristic of young objects. All targets observed have been well matched to the template spectra of young stellar objects, and therefore, we did not reject, in a first step, any candidate as a contaminant. Eight of the targets (CFHT-IC348-1, 4, 12, 14, 15, 22, 26, 29) show mid-IR excess consistent with the existence of a disc, hence we classify them as young members. These include two L0 dwarfs, CFHT-IC348-22 and CFHT-IC348-29, that had been previously identified as candidate members of the cluster based on their mid-IR excess by \citet[][SEDs are shown in Fig. 4 of their paper, IDs~1379 and 22865, respectively]{Muench2007}, but lacked spectroscopic confirmation. CFHT-IC348-22 also shows strong H$_{\alpha}$ emission. 

For the remaining 14 targets that do not show an indication of a disk, we must use other diagnostics. However, the low resolution and modest signal-to-noise of both optical and IR spectra, hinder the use of more accurate diagnostics of youth, such as the Na doublet in the optical, or the Na~I absorption line in the \emph{K} band. Although in the near-IR, the triangular shape of the \emph{H}-band caused by water absorption, with a peak redwards of that of field dwarfs  \citep{Luhman1999,Lucas2001,McGovern2004}, is one of the youth indicators, at low signal-to-noise particular combinations of extinction and spectral type make the numerical fitting converge for both young and field dwarfs. We have also crossed-matched our candidates with the new reduction of deep X-ray observations of IC 348 \citep{Stelzer2012} but none of our candidates has been detected, though only 2 of the previously known brown dwarfs were identified.

\begin{figure}
\centering
\includegraphics[width=\columnwidth]{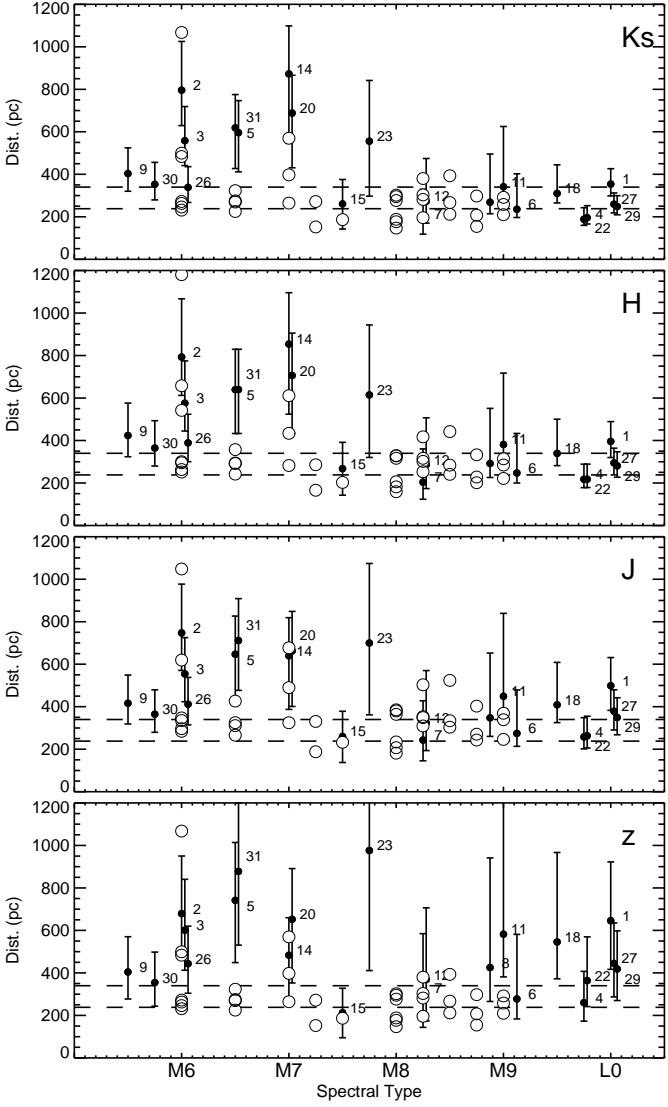}
\caption{Distance estimate for candidate members, by comparing their temperatures and magnitudes to those predicted by evolutionary models for young stellar objects (3~Myr). Objects with the same spectral type have been slightly shifted in the x-axis for clearness.}
\label{distance}
\end{figure}

Given the depth of our spectroscopic survey, the sources of contamination are field dwarfs in the foreground and background of the cluster, as well as background giants. We attempt at flagging those by comparing the observed properties to those predicted by the Dusty evolutionary models of young stellar objects \citep{Chabrier2000}. First, assuming all objects are young, we convert spectral types to temperatures adopting the temperature scale from \citet{Luhman2003ic348}. Secondly, we compute the distance at which the object would have to be to match the absolute magnitudes predicted by the 3~Myr old isochrones for its temperature. We have used the mean spectral type and extinction values for the targets with both optical and near-IR spectroscopy. The results are shown in Fig.~\ref{distance}, using the different photometric bands. Targets 2, 3, 5, 14, 20, and 31 appear to be at significantly higher distances than the IC 348 cluster ($\sim$240-340~pc), though not by more than $\sim$1.5~$\sigma$. This test is, however, very dependent on the accuracy of models, which have large uncertainties at this young ages, specially around spectral type M7 \citep{Dobbie2002}, where most of the outliers reside. 

\begin{figure}
\centering
\includegraphics[width=\columnwidth]{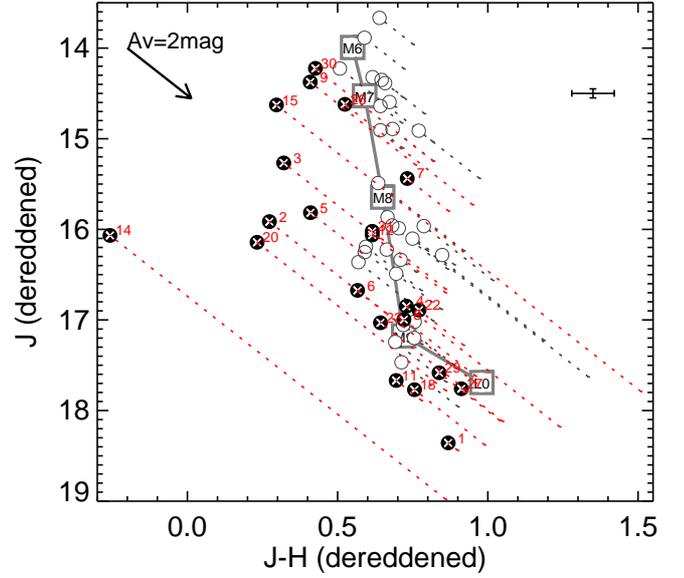}
\caption{Colour-magnitude diagram for the previously known brown dwarfs from the literature and the new candidate members corrected for extinction. The solid line represents the 3~Myr isochrone at the distance of the cluster, obtained by combining empirical relations of young stellar objects, and the Dusty models of \citet{Chabrier2000}. The dotted lines indicating the amount of extinction applied to each target. The typical error bar is shown.}
\label{cmddist}
\end{figure}

To further investigate the membership of the targets,  we compare the dereddened targets with the empirical isochrone for IC~348 in the \emph{J}~vs.~\emph{J}-\emph{H} colour-magnitude diagram constructed analogously to the method used by \citet{Luhman2003taurus}. This is done by converting the luminosity from the 3~Myr isochrone of the Dusty model to apparent magnitude, using an empirical temperature scale \citep{Luhman2003ic348} and bolometric corrections \citep{Dahn2002}, and a distance of 300~pc. The \emph{J}-\emph{H} colour is also empirically determined for young objects \citep{Luhman2010}. The results are shown in Fig.~\ref{cmddist}, where the known brown dwarfs from the literature (open circles) and the spectroscopically observed new candidate members (other circles), are shown dereddened, with a vector dotted line indicating the amount of extinction. The vast majority of the points are clustered around the isochrone, as expected for members of IC~348. Candidates CFHT-IC348-2, 3, 5, 14, 15, and 20 appear significantly bluer than the colours expected for young brown dwarfs. Candidate CFHT-IC348-15 has, however, strong mid-IR excess that could affect the near-IR colours and the extinction determination and is also a clear youth indicator, so we include it as a member. We classify the membership status of the other targets (CFHT-IC348-2, 3, 5, 14, and 20) as uncertain, and do not include them in the following analysis of the cluster. Furthermore, CFHT-IC348-31 is also too faint for the derived spectral type, and it is not even detected at mid-IR wavelengths, so it is classified as an uncertain member. The 16 remaining targets with spectroscopy are consistent with being at the distance of the cluster. That, together with their position in the various colour-magnitude diagrams, and the fact that their spectra are well matched by those of young stellar objects templates, are the criteria used to classified them as new members of IC~348. 

\subsection{The H-R diagram}\label{sectionhr}

\begin{figure}
\centering
\includegraphics[width=\columnwidth]{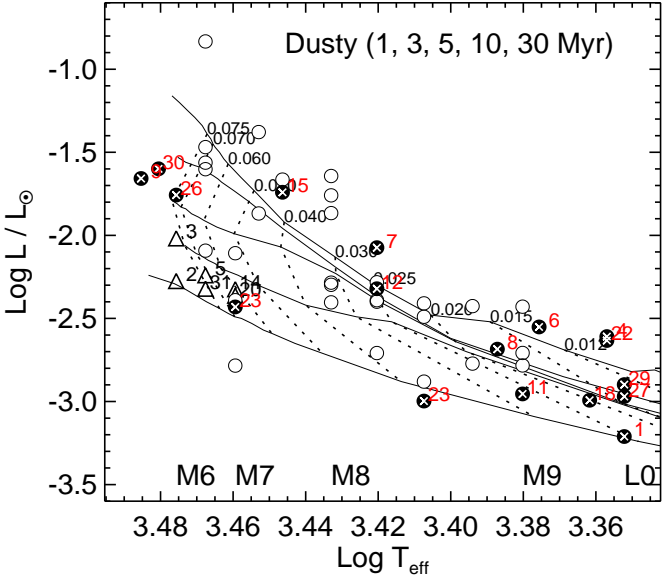}
\caption{H-R diagram for the new members of IC~348 spectroscopically confirmed (filled circles with white crosses), as well as the previously known brown dwarf members of the cluster (open circles). The models are the Dusty isochrones \citep{Chabrier2000} for 1, 3, 5, 10, and 30 Myr, labelled with mass in units of M$_{\sun}$. Additionally, candidates classified as having an uncertain membership status are shown as triangles. The cross denotes a typical error bar. }
\label{hr}
\end{figure}

To determine the masses of the new confirmed members of IC~348, we compare the effective temperatures and bolometric luminosities to the Dusty evolutionary models \citep{Chabrier2000} in the H-R diagram. We convert spectral types to temperatures adopting the temperature scale from \citet{Luhman2003ic348}. The bolometric luminosity was calculated using the dereddened \emph{J} magnitude, a distance to the cloud of 300~pc, and bolometric corrections from \citet{Dahn2002}. Figure~\ref{hr} shows all the new members overplotted with Dusty isochrones for 1, 3, 5, 10, and 30~Myr, as well as the previously known members from the literature with spectral type later than M6. Candidate CFHT-IC348-23 appears twice in the diagram showing the temperature and luminosity for both solutions from the optical and near-IR spectra, since it was the only target where the differences were significantly larger than the typical error of the numerical fitting procedure. Also shown are the candidates that we previously classified as having an uncertain membership status, that are seen clustered at relatively lower luminosity values than most of the cluster members. According to the evolutionary models, 13 sources (CFHT-IC348-1, 4, 6, 7, 8, 11, 12, 15, 18, 22, 23, 27, 29) have masses below the substellar limit, thus we classified them as new brown dwarf members of IC~348.

\section{The IMF}\label{sectionimf}
With the 13 new brown dwarfs found in this study, we can extend the substellar IMF of IC~348 to lower masses. We derived masses for the spectroscopically confirmed members from the 3~Myr evolutionary models, according to each target's effective temperature. We excluded from the analysis the members with uncertain membership status, and close binaries that are not resolved in the MegaCam and WIRCam images are treated as single objects. To ensure the derived properties are representative of the entire population (assuming that stars and brown dwarfs have the same spatial distribution), we defined an extinction limited sample, further restricted to the MegaCam and WIRCam completeness limits. This sample contains all objects laying bluewards of the 3 Myr Dusty isochrone reddened by 4 magnitudes of A$_{V}$, and brighter than the completeness limits in the \emph{z$^{\prime}$}~vs.~\emph{z$^{\prime}$}$-$\emph{J} diagram, which corresponds to a mass completeness limit of $\sim$13~M$_{\emph{Jup}}$ over the entire field of view of the survey. This completeness limited sample, contains 26 brown dwarfs from the literature, and 9 new brown dwarfs found in our study, all confirmed spectroscopically. 

In order to compare the new found objects with the stellar population of the cluster, we define another sample limited to the 16$\arcmin$~$\times$~14$\arcmin$ field from \citet[][complete from higher masses down to $\sim$30~M$_{\emph{Jup}}$]{Luhman2003ic348} to extend the IC~348 IMF down to $\sim$13~M$_{\emph{Jup}}$. Within this field of view and extinction-limited sample, our survey has resulted in only 3 additional sources. Two of these, CFHT-IC348-23 and 26, should have been detected by \citet{Luhman2003ic348}, though the authors state that a small percentage of their candidates had not been followed-up spectroscopically, but do not provide their coordinates. The other target, CFHT-IC348-11 has a mass below their completeness limit. 

\begin{figure}
\centering
\includegraphics[width=\columnwidth]{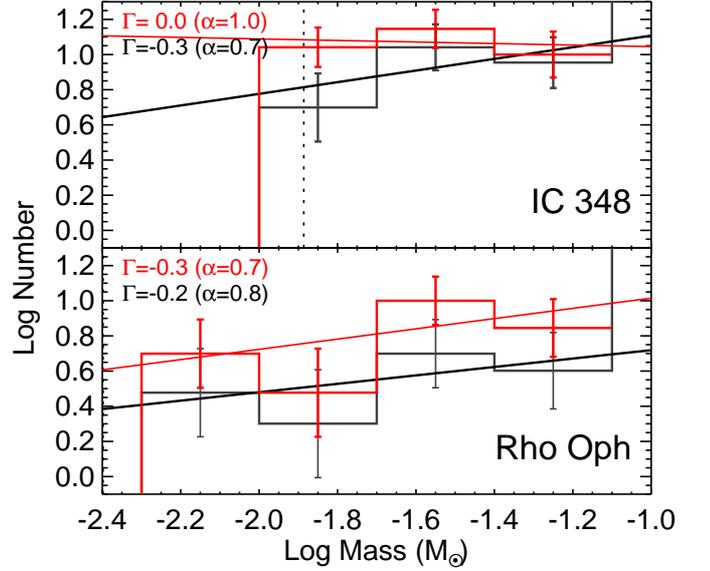}
\caption{\emph{Upper panel:} Substellar IMF for the spectroscopically confirmed population of substellar members of the IC~348 cluster for a complete and extinction-limited sample of A$_{V}$$\le$4~mag (red), and for the 16$\arcmin$~$\times$~14$\arcmin$ field from \citet{Luhman2003ic348} (black). The dotted line corresponds to the completeness limit of our survey. \emph{Lower panel:} Substellar IMF for the $\rho$~Ophiuchi cluster \citep{AlvesdeOliveira2012}, for two extinction-limited samples of A$_{V}$$\le$8~mag (black) and A$_{V}$$\le$15~mag (red). The best fit to a power law for masses $\le$80~M$_{\emph{Jup}}$ and the respective logarithmic exponent are shown$^{2}$.}
\label{imf}
\end{figure}

The resulting substellar IMFs for these two samples (one extinction-limited, the other additionally also limited to a smaller field-of-view) are shown in Fig.~\ref{imf}. For comparison, we also include the substellar IMF of $\rho$~Oph \citep{AlvesdeOliveira2012} for two extinction limited samples of A$_{V}$=8~mag and A$_{V}$=15~mag, complete down to $\sim$4~M$_{\emph{Jup}}$. The best fit\footnote{Note that $\alpha$=$\Gamma$+1.} to a power-law for masses $\le$80~M$_{\emph{Jup}}$ gives an $\alpha$ of 1$\pm$0.3 and 0.7$\pm$0.4 for the IC~348 substellar IMF extinction-limited, and spatially limited to the centre of the cluster, respectively. We note that these results do not change significantly if we include the 5 targets with the uncertain membership, with the $\alpha$ values agreeing within the errors. For $\rho$~Oph, we find an $\alpha$ of 0.8$\pm$0.4 for an extinction limited sample for A$_{V}$$\le$8~mag, and 0.7$\pm$0.3 for the entire WIRCam field complete to A$_{V}$$\le$15~mag in the substellar regime. 

\begin{figure}
\centering
\includegraphics[width=\columnwidth]{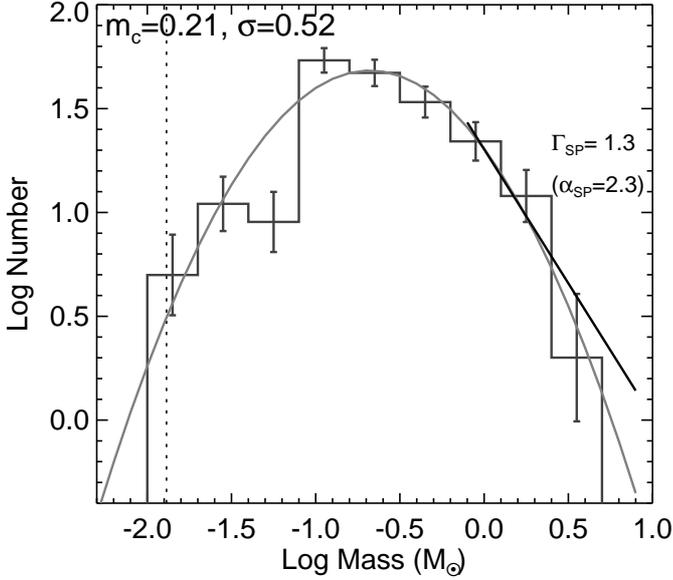}
\caption{Spectroscopic IMF for IC~348 for a complete and extinction-limited sample of A$_{V}$$\le$4~mag. The best-fit to a lognormal is shown, as well as the best-fit high-mass slope (which matches the Salpeter slope). }
\label{imfall}
\end{figure}

\begin{figure}
\centering
\includegraphics[width=\columnwidth]{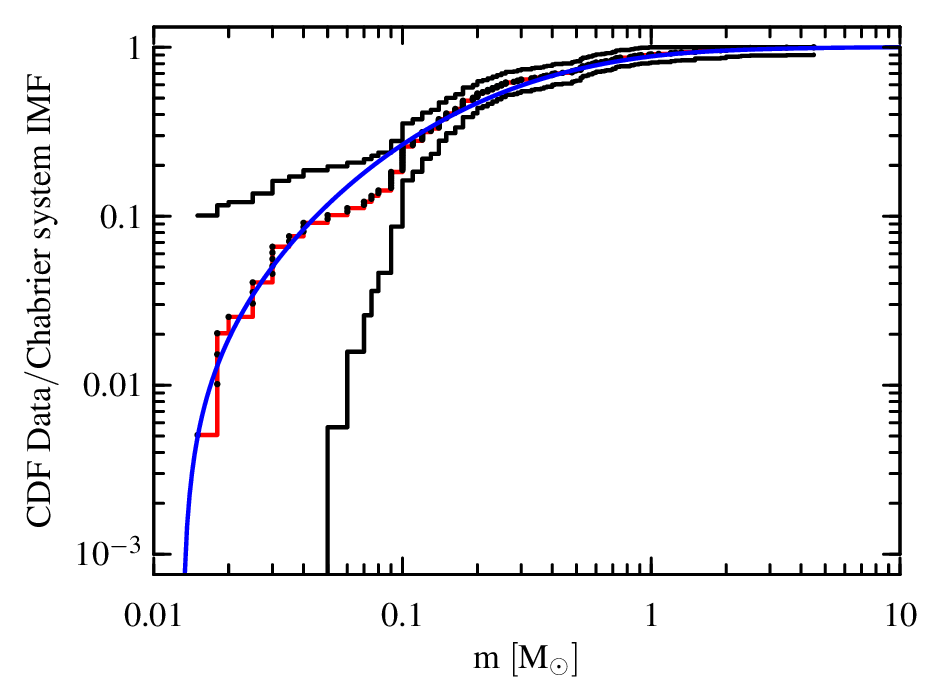}
\caption{Cumulative distribution function for the complete and extinction-limited sample of IC~348 members (red). The \citet{Chabrier2003} CDF is overplotted (blue), as well as the 95\% confidence intervals for a KS test (black).}
\label{imfall2}
\end{figure}

Figure~\ref{imfall} shows the IMF of IC~348 across all masses, as determined by \citet{Luhman2003ic348}, with the new members spectroscopically confirmed in our study added. We have fitted the data to a log-normal distribution and find the best fit for a characteristic mass of m$_{c}$$=$0.21$\pm$0.02~M$_{\sun}$ and $\sigma$$=$0.52$\pm$0.03~dex. For higher masses we recover the Salpeter slope with the best fit result of $\alpha_{SP}$$=$2.3$\pm$0.4. We have further analyzed the cumulative distribution function (CDF) of the masses in the same complete and extinction-limited sample, and compared it to the \citet{Chabrier2003} system IMF (m$_{c}$$=$0.22~M$_{\sun}$ and $\sigma$$=$0.57~dex). Figure~\ref{imfall2} shows the IC~348 CDF, with the Chabrier IMF overplotted in blue, and the 95\% confidence intervals for a KS test shown in black. There is no significant deviation of the data from the Chabrier-type hypothesis.

The parameters derived for the best fit to the log-normal IMF \citep{Chabrier2003} are in agreement with the IMFs derived for other clusters of similar ages (see, for example, the fit of the fiducial lognormal mass function of the Pleiades from \citet{Moraux2003} to young clusters in Fig.~17 from \citet{Bayo2011}). Similarly, when comparing the slope of the power law in the substellar regime with other young clusters, as for example our previous work in $\rho$~Oph shown in Fig.~\ref{imf}, the spectroscopic IMFs in NGC~1333 \citep[][]{Scholz2012}, Upper Sco \citep{Lodieu2007,Lodieu2008}, Collinder 69 \citep{Bayo2011}, or the photometric IMF in $\sigma$~Ori \citep[][and references therein for other clusters]{Bejar2011}, we find it to be consistent with other regions where $\alpha$ measurements vary between $\sim$0.3 and 1.0. This suggests that also in IC~348, the substellar mass function does not show variations from other young clusters, down to the planetary-mass limit.

An extrapolation of the log-normal distribution estimates only 1 object in the mass range 5-10~M$_{\emph{Jup}}$, and $<$1 object below 5~M$_{\emph{Jup}}$. In a complementary study of IC~348 within the WIRCam survey using methane imaging to search for T dwarf members of few Jupiter masses, \citet{Burgess2009} found only one suitable candidate, which is still awaiting spectroscopic confirmation. If confirmed as a member, this result would be consistent with the predictions from extrapolation of the log-normal IMF. However, we cannot rule out a mass cut-off below the completeness limits of our survey. Deeper photometric and spectroscopic studies, as well as an improvement in the reliability of mass determination from evolutionary models, are needed to understand the mass distribution in the planetary mass regime.

\section{Conclusions}

A deep near-IR and optical photometric survey of the young cluster IC~348 was conducted with MegaCam and WIRCam at the CFHT to search for low-mass brown dwarfs members of the cluster. We identified 31 candidate members with no previous spectroscopic information. The follow-up of 24 candidates with optical and/or near-IR spectroscopy resulted in the confirmation of 16 new members of the cluster. Six targets have an uncertain membership, and the spectra of 2 targets did not have enough signal-to-noise, and therefore those remain as possible members. 

According to evolutionary models, 13 of the new spectroscopically confirmed members have masses below the substellar limit, and we classified them as new brown dwarfs members of IC~348. Amongst the new members of the cluster are 5 brown dwarfs with spectral type L0, the latest-type objects confirmed to date in IC~348. At 3~Myr, evolutionary models estimate a mass of 13~M$_{\emph{Jup}}$ for these objects. Two of the L0 dwarfs (CFHT-IC348-22 and 29) have mid-IR excess indicative of a circumstellar disc. 

We combine the newly found low-mass brown dwarfs with previous census of the cluster and construct the IMF, complete down to $\sim$13~M$_{\emph{Jup}}$ and A$_{V}$$\le$4~mag. We find the IC~348 mass function to be well fitted by the log-normal \citet{Chabrier2003} system IMF. Furthermore, we do not find evidence for a variation of the IMF of IC~348 when compared to other young clusters, down to the planetary-mass regime.


\begin{acknowledgements}
We thank the QSO team at CFHT for their efficient work at the telescope and the data pre-reduction as well as the Terapix group at IAP for the image reduction. We thank the anonymous referee for the useful suggestions. This work is based in part on data products produced and image reduction processes conducted at TERAPIX. Based on observations obtained at the Gemini Observatory, which is operated by the Association of Universities for Research in Astronomy, Inc., under a cooperative agreement with the NSF on behalf of the Gemini partnership: the National Science Foundation (United States), the Science and Technology Facilities Council (United Kingdom), the National Research Council (Canada), CONICYT (Chile), the Australian Research Council (Australia), Minist\'erio da Ci\^encia, Tecnologia e Inova‹o (Brazil) and Ministerio de Ciencia, Tecnolog\'{\i}a e Innovaci\'on Productiva (Argentina). Based on observations made with the Gran Telescopio Canarias (GTC), instaled in the Spanish Observatorio del Roque de los Muchachos of the Instituto de Astrof'sica de Canarias, in the island of La Palma. This publication makes use of data products from the Wide-field Infrared Survey Explorer, which is a joint project of the University of California, Los Angeles, and the Jet Propulsion Laboratory/California Institute of Technology, funded by the National Aeronautics and Space Administration. This research has made use of the NASA/ IPAC Infrared Science Archive, which is operated by the Jet Propulsion Laboratory, California Institute of Technology, under contract with the National Aeronautics and Space Administration. This research has made use of the SIMBAD database, operated at CDS, Strasbourg, France. Research was partly supported by the French National Research Agency under grant 2010 JCJC 0501 1.
\end{acknowledgements}


\bibliographystyle{aa}
\bibliography{ic348}

\begin{landscape}
   \begin{table}   
\centering             
   \tiny
   \caption{Candidate members of IC~348.}           
\begin{tabular}{@{ }l@{ }l@{ }l@{ }l@{ }l@{ }l@{ }l@{ }l@{ }l@{ }l@{ }l@{ }l@{ }l@{ }l@{ }l@{ }l@{ }l@{ }l@{ }l@{ }}
   \hline      \hline

CFHT-IC348 & R.A.     & Dec. & \emph{z$^{\prime}$}\tablefootmark{a} & \emph{J}\tablefootmark{b} & \emph{H}\tablefootmark{b} & \emph{K$_{s}$}\tablefootmark{b} &  $[3.6]$\tablefootmark{c} & $[4.5]$\tablefootmark{c} & $[5.4]$\tablefootmark{c} & $[8.0]$\tablefootmark{c} &  Properties\tablefootmark{d} & Sp.T.\tablefootmark{e} & A$_{v}$\tablefootmark{e} & Sp.T. \tablefootmark{e} & A$_{v}$\tablefootmark{e} & L$_{bol}$\tablefootmark{f} & M\tablefootmark{f} & Membership  \\
    & (J2000) &  (J2000) & (mag)   & (mag)    & (mag)  & (mag)   & (mag) & (mag) & (mag) & (mag) &    &  \multicolumn{2}{c}{Optical} & \multicolumn{2}{c}{Near-IR} &  (L$_{\sun}$)  & (M$_{\sun}$) &     \\
    &   &   & ($\pm$0.05) & ($\pm$0.05)  & ($\pm$0.05)  & ($\pm$0.05)  &    &    &    &    &	 &  & & & & & &        \\
\hline
1 	  &  03:43:22.51  &  +32:18:17.5  &  21.06  &  18.44  &  17.54  &  16.85 &  15.79$\pm$0.08 & 15.48$\pm$0.09 &	   	   	 &		&	 s,ex2  	&		&		&   L0  	&   0.3 	    &  0.0006   &     0.013	 & spectroscopic member     \\      
2 	  &  03:43:25.71  &  +32:21:54.2  &  19.08  &  17.08  &  16.36  &  15.87 &  15.28$\pm$0.08 & 15.26$\pm$0.09 &	   	   	 &		&	 s		&  M6		&  4.2  	&		&		    &	       &		 & uncertain		    \\      
3 	  &  03:43:31.38  &  +32:00:19.8  &  18.19  &  16.10  &  15.46  &  14.96 &  14.39$\pm$0.06 & 14.30$\pm$0.06 &  13.99$\pm$0.11  &      	      &        s	      &  M6	      &  3.0	      & 	      & 		  &	     &  	       & uncertain		  \\  	  
4 	  &  03:43:42.41  &  +32:09:32.7  &  21.31  &  18.18  &  16.94  &  15.94 &  14.96$\pm$0.06 & 14.47$\pm$0.07 &  14.21$\pm$0.23  & 13.45$\pm$0.43 &	 s,ex1,ex2	&		&		&   M9.75	&   4.8 	    &  0.0025   &     0.014	 & spectroscopic member     \\  	       
5 	  &  03:43:45.36  &  +31:55:25.9  &  18.17  &  16.26  &  15.68  &  15.23 &  14.72$\pm$0.06 & 14.56$\pm$0.06 &  14.67$\pm$0.16  &               &	s	       &  M6.5         &  1.6	       &	       &		   &	      & 		& uncertain		   \\       
6 	  &  03:44:07.27  &  +32:14:33.2  &  20.42  &  17.70  &  16.74  &  16.04 &  15.19$\pm$0.07 & 15.02$\pm$0.08 &	    	   &      	       &	s	       &  L0	       &  2.7	       &   M8.25       &   4.7  	   &  0.0028   &     0.016	& spectroscopic  member    \\      
7 	  &  03:44:11.51  &  +32:18:09.6  &  18.06  &  15.80  &  14.93  &  14.35 &  13.78$\pm$0.06 & 13.66$\pm$0.06 &  13.57$\pm$0.10  & 13.63$\pm$0.25 &	 s		&  M8.25	&  1.3  	&		&		    & 0.0084   &     0.025	 & spectroscopic member     \\  	       
8 	  &  03:44:15.47  &  +32:19:23.1  &  20.31  &  17.62  &  16.66  &  15.98 &  15.31$\pm$0.07 & 15.16$\pm$0.08 &  15.11$\pm$0.35  &      		&	 s		&  M8.25	&  2.4  	& M9.5  	&	2.1	    & 0.0021   &     0.018	 & spectroscopic member     \\      
9 	  &  03:44:18.08  &  +31:57:00.0  &  17.17  &  15.29  &  14.53  &  14.02 &  13.56$\pm$0.05 & 13.43$\pm$0.06 &  13.34$\pm$0.07  & 13.16$\pm$0.11 &	 s		&  M5.5 	&  3.3  	&		&		    & 0.0220   &     0.109	 & spectroscopic member     \\   	       
10	  &  03:44:18.50  &  +32:19:35.1  &  22.03  &  19.11  &  18.10  &  17.39 &  16.51$\pm$0.12 & 16.61$\pm$0.19 &	    	     &      		&	 s		&		&		&   poor S/N	&  poor S/N	    &	       &		 & photometric candidate    \\      
11	  &  03:44:18.76  &  +32:11:32.0  &  20.41  &  17.89  &  17.11  &  16.44 & 	           &	            &	    	       &           	&	 s		&		&		& M9		&	0.8	    &  0.0011   &     0.017	 & spectroscopic member     \\      
12	  &  03:44:18.85  &  +32:19:32.8  &  18.99  &  16.70  &  15.84  &  15.26 &  14.67$\pm$0.06 & 14.44$\pm$0.07 &  14.30$\pm$0.20  & 13.54$\pm$0.43 &	 s,ex1,ex2	&  M8.25	&  2.3  	&		&		    &  0.0048   &     0.025	 & spectroscopic member     \\  	       
13	  &  03:44:25.93  &  +32:08:05.4  &  22.17  &  19.05  &  18.07  &  17.34 &  16.60$\pm$0.26 & 16.46$\pm$0.26 &	    	    	  &		    &	     s  	    &		    &		    &	poor S/N    &	poor S/N	&	   &		     & photometric candidate	\\  	
14	  &  03:44:34.54  &  +32:02:13.7  &  22.18  &  18.98  &  18.12  &  17.29 &  16.45$\pm$0.08 & 16.18$\pm$0.13 &	    	    	  &		    &	     s,ex2	    &		    &		    &	M7	    &	10.5		&	   &		     & uncertain		\\  	
15	  &  03:44:39.28  &  +32:14:43.0  &  21.43  &  17.79  &  16.28  &  15.28 &  14.31$\pm$0.06 & 13.94$\pm$0.06 &  13.99$\pm$0.19  & 12.91$\pm$0.29 &	 s,ex1,ex2	&		&		& M7.5  	&11.4		    &  0.0182  &     0.040	 & spectroscopic member     \\         
16	  &  03:44:40.98  &  +32:01:49.9  &  21.17  &  17.70  &  16.22  &  15.26 &  14.61$\pm$0.06 & 14.45$\pm$0.06 &  14.33$\pm$0.12  &               &		       &	       &	       &	       &		   &	      & 		& photometric candidate    \\       
17	  &  03:44:49.09  &  +31:58:51.0  &  19.88  &  17.40  &  16.37  &  15.75 &  15.34$\pm$0.06 & 15.20$\pm$0.07 &  14.69$\pm$0.26  &      	       &		       &	       &	       &	       &		   &	      & 		& photometric candidate    \\       
18	  &  03:44:49.33  &  +32:09:49.5  &  21.45  &  18.38  &  17.39  &  16.66 &  15.84$\pm$0.11 & 15.87$\pm$0.19 &	   	   	 &		 &	  s		 &		 &		 &	 M9.5	 &	 2.2	     &  0.0010	&     0.014	  & spectroscopic member     \\      
19	  &  03:44:50.25  &  +32:06:35.6  &  21.35  &  18.63  &  17.82  &  17.21 &  16.33$\pm$0.09 & 16.08$\pm$0.12 &	   	   	 & 12.87$\pm$0.48 &	   ex2  	  &		  &		  &		  &		      & 	 &		   & photometric candidate    \\      
20	  &  03:44:50.52  &  +31:56:57.3  &  20.44  &  17.78  &  16.92  &  16.24 &  15.56$\pm$0.06 & 15.40$\pm$0.07 &  15.13$\pm$0.20  &              	  &	   s		  &		  &		  & M7  	  & 5.9 	      & 	 &		   & uncertain  	      \\       
21	  &  03:44:51.48  &  +32:19:54.0  &  18.36  &  16.10  &  15.20  &  14.62 &  14.06$\pm$0.07 & 13.93$\pm$0.07 &	    	    &      	      	  &			  &		  &		  &		  &		      & 	 &		   & photometric candidate    \\      
22	  &  03:44:52.00  &  +31:59:21.6  &  20.27  &  17.28  &  16.36  &  15.63 &  14.92$\pm$0.06 & 14.45$\pm$0.06 &  13.98$\pm$0.08  & 13.32$\pm$0.11 &	 s,ex1,ex2	&  M9.5 	&  0.8  	&   L0  	&   2.0 	    & 0.0023  &     0.014	 & spectroscopic member     \\       
23	  &  03:44:57.34  &  +32:08:35.1  &  20.50  &  17.78  &  16.85  &  16.16 &  15.51$\pm$0.07 & 15.31$\pm$0.09 &	    	    &      		&	 s		&  M8.5 	&  0.1  	& M7		&	5.3	    & 0.0024   &     0.035	 & spectroscopic member     \\      
24	  &  03:45:03.78  &  +32:21:34.5  &  15.65  &  13.95  &  13.30  &  12.77 &  12.41$\pm$0.06 & 12.17$\pm$0.06 &  12.11$\pm$0.07  & 12.02$\pm$0.08 &			&		&		&		&		    &	       &		 & photometric candidate    \\      
25	  &  03:45:03.80  &  +32:24:22.5  &  16.78  &  14.57  &  13.69  &  13.12 &  12.64$\pm$0.06 & 12.47$\pm$0.06 &  12.29$\pm$0.07  & 12.22$\pm$0.08 &			&		&		&		&		    &	       &		 & photometric candidate    \\      
26	  &  03:45:06.44  &  +31:59:39.4  &  18.05  &  15.73  &  14.78  &  13.99 &  13.47$\pm$0.06 & 13.20$\pm$0.05 &  12.99$\pm$0.07  & 12.58$\pm$0.07 &	 s,ex1,ex2	&  M6		&  4.0  	&		&		    & 0.0175   &     0.086	 & spectroscopic member     \\       
27	  &  03:45:08.87  &  +32:15:46.9  &  20.77  &  18.12  &  17.07  &  16.29 &  15.48$\pm$0.07 & 15.28$\pm$0.09 &  15.19$\pm$0.36  &              	 &	  s		 &		 &		 &  L0  	 &   1.3	     & 0.0011	&     0.013	  & spectroscopic member     \\  
28	  &  03:45:12.84  &  +32:03:36.0  &  20.23  &  17.89  &  17.13  &  16.43 &  15.65$\pm$0.06 & 15.48$\pm$0.09 &	    	     &      	      	 &			 &		 &		 &		 &		     &  	&		  & photometric candidate    \\      
29	  &  03:45:17.65  &  +32:07:55.3  &  20.95  &  18.11  &  17.07  &  16.27 &  15.29$\pm$0.06 & 14.90$\pm$0.07 &  14.88$\pm$0.21  &      	      	 &	  s,ex2 	 &		 &		 & L0		 &	 1.9	     & 0.0013 	&     0.013	  & spectroscopic member     \\      
30	  &  03:45:23.90  &  +32:11:54.5  &  17.25  &  15.25  &  14.43  &  13.91 &  13.41$\pm$0.06 & 13.21$\pm$0.06 &  13.22$\pm$0.09  & 13.05$\pm$0.14 &	 s		&  M5.75	&  3.7  	&		&		    & 0.0251    &     0.097	 & spectroscopic member     \\      
31	  &  03:45:31.45  &  +32:12:28.9  &  18.90  &  16.66  &  15.80  &  15.23 & 	           & 14.47$\pm$0.07 &	    	       & 14.06$\pm$0.29 &	 s		&  M6.5 	&  2.3  	&		&		    &      &      	 & uncertain     \\      
\hline
\end{tabular}
\label{candidates} 

\tablefoottext{a}{PSF photometry magnitude from MegaCam / CFHT.} 
\tablefoottext{b}{PSF photometry magnitudes from WIRCam / CFHT.} 
\tablefoottext{c}{IRAC / Spitzer data as retrieved from the NASA/ IPAC Infrared Science Archive.}
\tablefoottext{d}{s~=~spectroscopic follow-up; ex1, ex2~=~mid-IR excess. See Sect.~\ref{select:cmd} for details.} 
\tablefoottext{e}{Spectral Type and A$_{\emph{V}}$ as determined from this study. See Sect.~\ref{mem} for details.}
\tablefoottext{f}{Luminosity and mass, as determined from evolutionary models. See Sects.~\ref{sectionhr} and \ref{sectionimf} for details.}

\end{table}
\end{landscape}

\end{document}